\documentclass{article}
\usepackage[preprint]{spconf}
\usepackage{cite}
\usepackage{amsmath,amssymb,amsfonts}
\usepackage{graphicx}
\usepackage{subcaption}
\usepackage{textcomp}
\usepackage{algorithm}
\usepackage{algpseudocode}
\usepackage{booktabs, multirow} 
\usepackage{soul}
\usepackage[table]{xcolor} 
\usepackage{changepage,threeparttable} 
\usepackage[hyphens]{url}
\usepackage{arydshln}
\usepackage{tabularx}
\usepackage{cprotect}
\def\BibTeX{{\rm B\kern-.05em{\sc i\kern-.025em b}\kern-.08em
    T\kern-.1667em\lower.7ex\hbox{E}\kern-.125emX}}

\title{Segment-Level Vectorized Beam Search Based on\\Partially Autoregressive Inference}

\name{
    \begin{tabular}{c}
    Masao Someki$^1$, 
    Nicholas Eng$^2$, 
    Yosuke Higuchi$^3$, 
    Shinji Watanabe$^4$
    \end{tabular}
}
\address{
    $^1$IBM Japan Ltd., Japan \quad
    $^2$The University of Auckland, New Zealand \\
    $^3$Waseda University, Japan \quad
    $^4$Carnegie Mellon University, USA
}
\copyrightnotice{979-8-3503-0689-7/23/\$31.00~\copyright2023 IEEE}
\begin{document}
\ninept

\maketitle

\begin{abstract}
Attention-based encoder-decoder models with autoregressive (AR) decoding have proven to be the dominant approach for automatic speech recognition (ASR) due to their superior accuracy.
However, they often suffer from slow inference.
This is primarily attributed to the incremental calculation of the decoder.
This work proposes a partially AR framework, which employs segment-level vectorized beam search for improving the inference speed of an ASR model based on the hybrid connectionist temporal classification (CTC) attention-based architecture.
It first generates an initial hypothesis using greedy CTC decoding, identifying low-confidence tokens based on their output probabilities.
We then utilize the decoder to perform segment-level vectorized beam search on these tokens, re-predicting in parallel with minimal decoder calculations.
Experimental results show that our method is $12$ to $13$ times faster in inference on the LibriSpeech corpus over AR decoding whilst preserving high accuracy.
\end{abstract}


\section{Introduction}

Due to recent advances in deep learning,
automatic speech recognition (ASR) has witnessed remarkable achievements~\cite{Hinton2012,graves2013,prabhavalkar2023end}.
ASR plays an essential role in facilitating human-computer interaction by providing an interface for converting audio to text and
demonstrating substantial applicability in real-world scenarios.
In particular, the RNN-Transducer model~\cite{rnn_transducer},
which operates relatively fast and can be extended to streaming speech recognition, is widely used in real-world applications.
Recent research in ASR has also made significant progress in achieving higher accuracy through Attention-based Encoder-Decoder (AED) models \cite{conformer,e_branchformer}.
AED models have also been utilized in models such as Whisper~\cite{whisper} and speech translation~\cite{inaguma-etal-2020-espnet}, and their usefulness has been reevaluated.
However, there are various trade-offs that can potentially limit its application in certain scenarios.
For example, one can construct an ASR system with high accuracy
through large and complex models,
but this comes at the expense of increased computational cost and inference time.

There has been extensive research devoted to alleviating the trade-off between recognition accuracy and inference speed.
Inspired by the great success in neural machine translation~\cite{gu2018non,xiao2023survey},
non-autoregressive (NAR) models have been actively studied in the context of ASR,
with the aim of achieving fast inference~\cite{chen2021non,chan2020imputer,higuchi20b_interspeech,chi2021align}.
Compared to the conventional autoregressive (AR) models~\cite{seq2seqNN,listenAttendSpell},
which generates output at each step conditioned on the previously generated outputs,
NAR models can produce multiple outputs simultaneously.
This parallel computation accelerates the inference process of ASR,
resulting in a significant reduction in inference time compared to AR models~\cite{Higuchi2021}.
However, it remains a challenge to achieve the same level of recognition accuracy as AR models.
In addition,
NAR models require a complex model structure or a unique training strategy
for the successful implementation of parallel generation during inference~\cite{Higuchi2021}.

Regarding AR decoding, the decoder is trained to learn linguistic information.
This allows us to utilize the relationship between the current token and the previous token.
Furthermore, it is common to enhance accuracy by implementing the beam search algorithm, which is a heuristic search for the best hypothesis.
However, because AR requires the previous tokens to estimate the next token,
it is not possible to parallelize the inference of a single audio.
As a consequence, achieving NAR-level speed through inference parallelization is challenging.

In this paper, we focus on the difference in trade-off balance between AR and NAR,
and propose a new decoding method, the partially autoregressive (PAR) method 
as a fast and accurate decoding method.
%
By utilizing the NAR approach and segment-level vectorized beam search, we can compensate for the weaknesses of both AR and NAR.
This results in a better trade-off between accuracy and latency.
In particular, we show that by optimizing the inference operations of the pre-trained hybrid connectionist temporal classification (CTC) and attention model,
we are able to achieve NAR-level inference speed while maintaining its high accuracy, as well as not needing additional training of the model.
In this paper, we make the following contributions:
\begin{itemize}
    \item We propose a new decoding framework that combines AR and NAR.
    \item We demonstrate a better balance of accuracy-latency trade-off without additional training.
\end{itemize}

\begin{figure*}[htb!]
    \centering
    \begin{subfigure}[b]{0.25\textwidth}
        \centering
        \includegraphics[width=\textwidth]{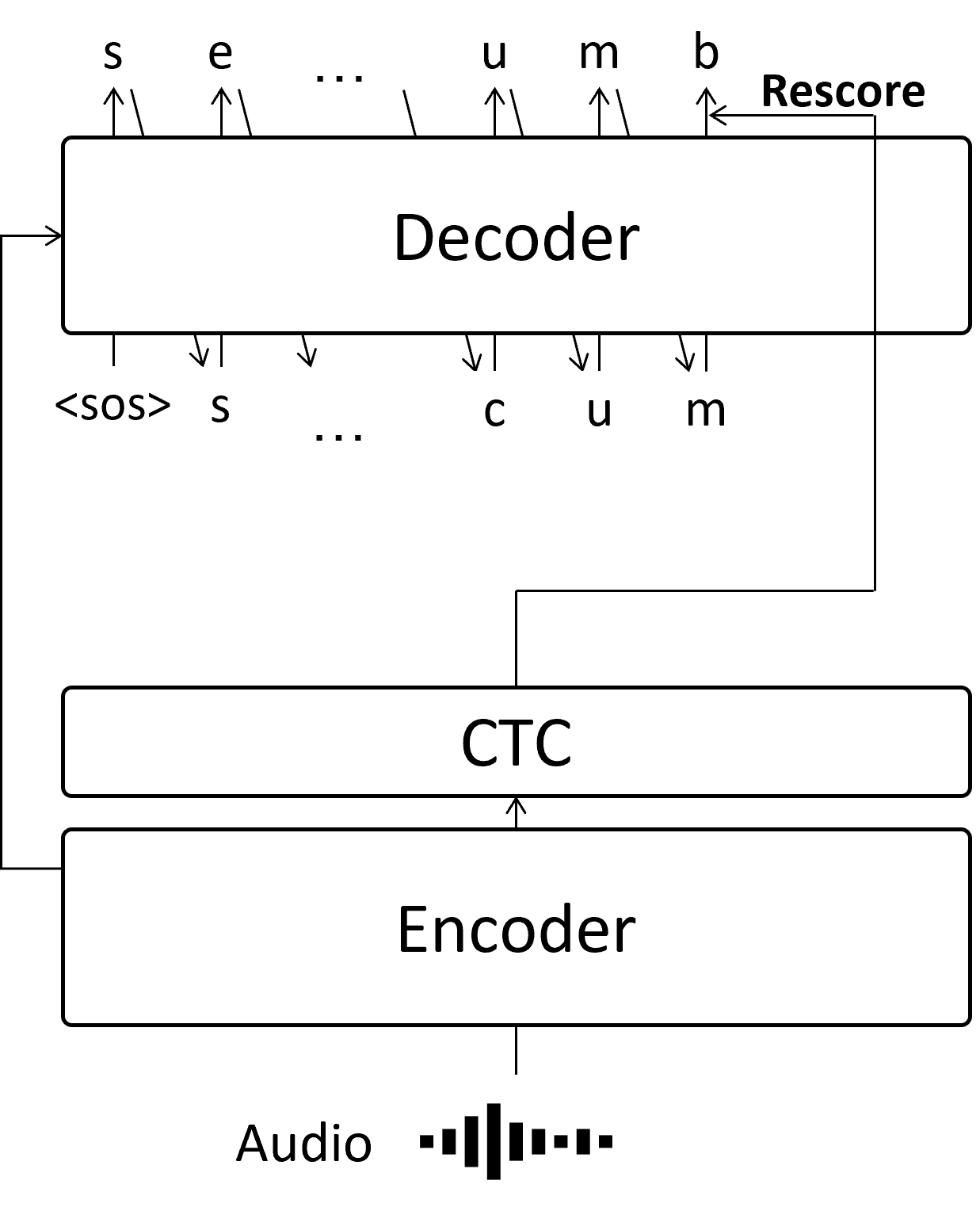}
        \caption{Autoregressive (AR)}
        \label{fig:ar_decode}
    \end{subfigure}
    \hspace{1.5cm}
    \begin{subfigure}[b]{0.25\textwidth}
        \centering
        \includegraphics[width=\textwidth]{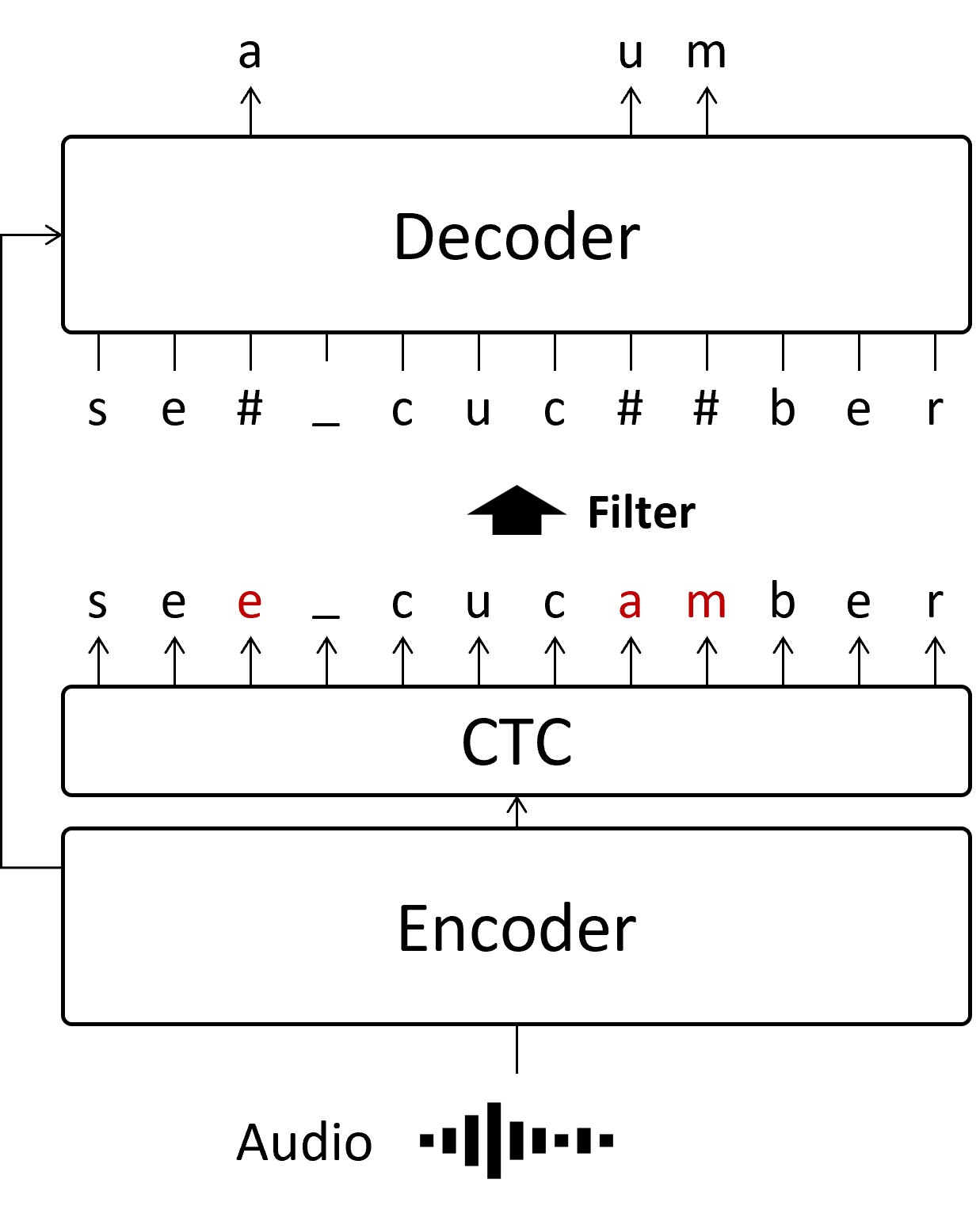}
        \caption{Non-autoregressive (NAR)}
        \label{fig:nar_decode}
    \end{subfigure}
    \hspace{1.5cm}
    \begin{subfigure}[b]{0.25\textwidth}
        \centering
        \includegraphics[width=\textwidth]{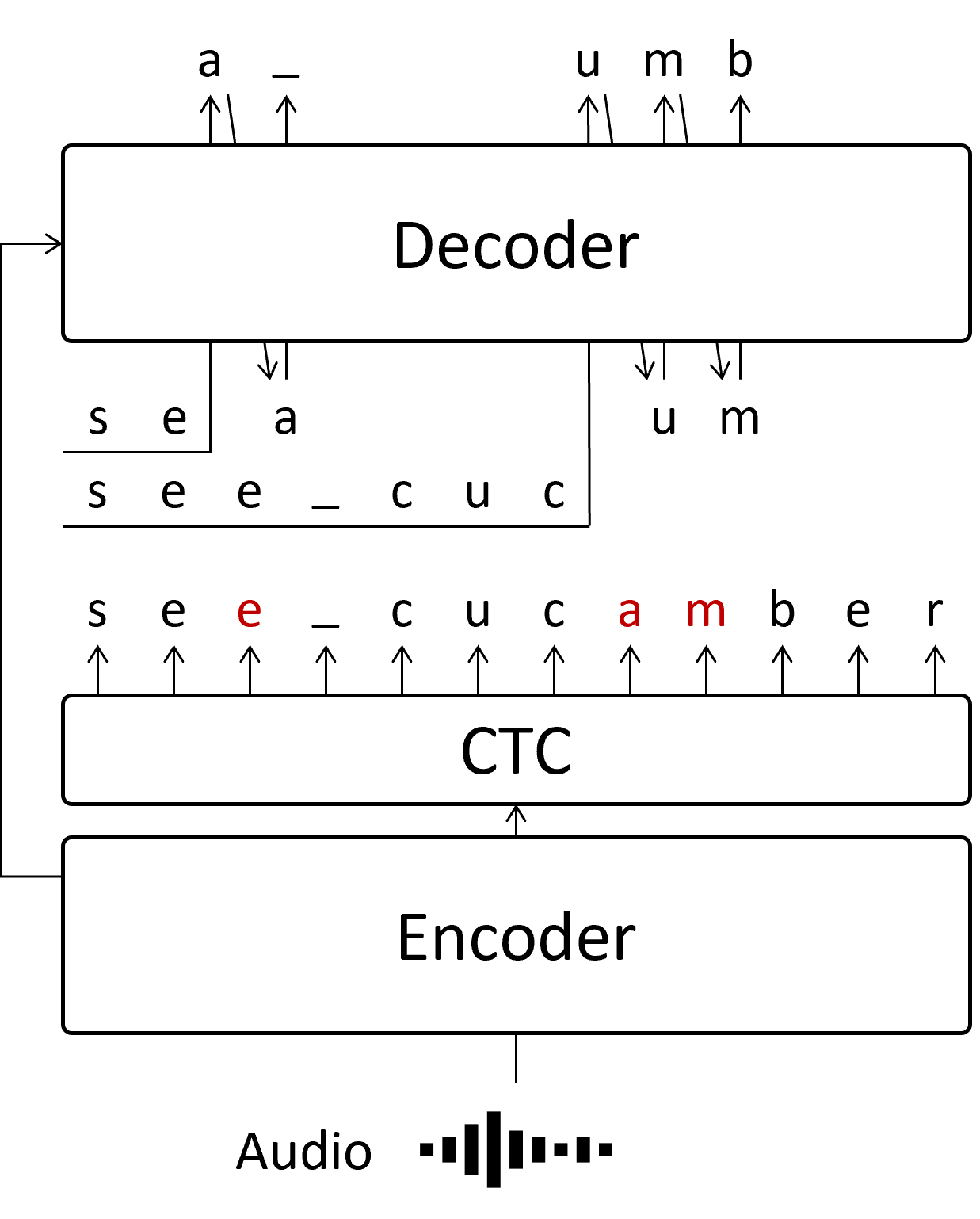}
        \caption{Partially autoregressive (PAR)}
        \label{fig:PAR_decode}
    \end{subfigure}
\cprotect\caption{Overview and comparison of AR, NAR, and PAR decoding. \verb|<sos>| denotes the ``start-of-sequence'' symbol,and the mask token is denoted by \verb|#| or red characters.
PAR is a hybrid of AR and NAR methods, in which the masking process is applied first, followed by segment-level vectorized beam search.}
\end{figure*}

\section{Related Works}

Numerous studies have been conducted to balance the accuracy-latency trade-off. 
In general, there are two main approaches;
reducing latency while preserving good accuracy or improving accuracy while preserving fast inference speed.

To reduce latency, there are several methods to lower the computational cost during inference.
One such method is the pruning technique~\cite{lecun1989optimal,frankle2019lth,Cheng2021perp}, 
which reduces the number of parameters in a trained model by identifying subnetworks with better performance and fewer parameters.
Another method is knowledge distillation~\cite{Hinton2015Distill,Chang2022distillhubert},
which trains a smaller model to reduce the number of parameters required during inference. 
To further reduce the number of search iterations, hypotheses can be predicted in a batch~\cite{Seki2019}, or the search process can be stopped prematurely~\cite{Jiahui2021fastemit}.
In addition, other approaches focus on utilizing machine resources more efficiently\cite{yao2021wenet,zhang2022wenet,masao2022espnetonnx},
rather than relying solely on the model architecture, to achieve faster processing.
In this study, we parallelized the AR decoding of single audio to achieve high-speed processing. 
The process is similar to batch processing of hypotheses, but in this case, the audio is segmented and parallelized in addition to the hypotheses.

On the other hand, we can also improve the accuracy of a low-latency model, such as a
non-autoregressive (NAR) model, though at a lower accuracy than typical AR models. 
For example, by replacing and inserting the model output with a mask sequence \cite{chen2021non,chan2020imputer,higuchi20b_interspeech, Higuchi2021maskctc}, NAR can more accurately predict the target sequence compared to a standard NAR model. 
We can also improve the accuracy by utilizing an external language model\cite{futami22_interspeech,higuchi2023bert}.
Additionally, several investigations have been undertaken to improve both the inference speed and accuracy, such as parallelizing the decoding process in the streaming situation\cite{Mahadeokar2022StreamingPT}.
In our work, it is not possible to perform parallel processing of all tokens as investigated above.
However, high accuracy can still be achieved by utilizing AR decoding in areas where parallel processing leads to decreased accuracy.

Research has investigated the combination of AR and NAR methods \cite{Zhengkun2022}.
The researchers trained a dual-mode Transformer decoder that can be used for both NAR and AR-style processes.
They first applied NAR-style decoding to generate several hypotheses, then used AR-style rescoring to select the best hypothesis.
Our approach indeed uses AR decoding after NAR decoding.
However, we utilize NAR decoding in our work to parallelize AR decoding within a single audio, so the purpose of NAR decoding is different.

\begin{figure}[!htp]
\begin{tabular}{c}
  \begin{minipage}[t]{\linewidth}
    \centering
        \includegraphics[width=\textwidth]{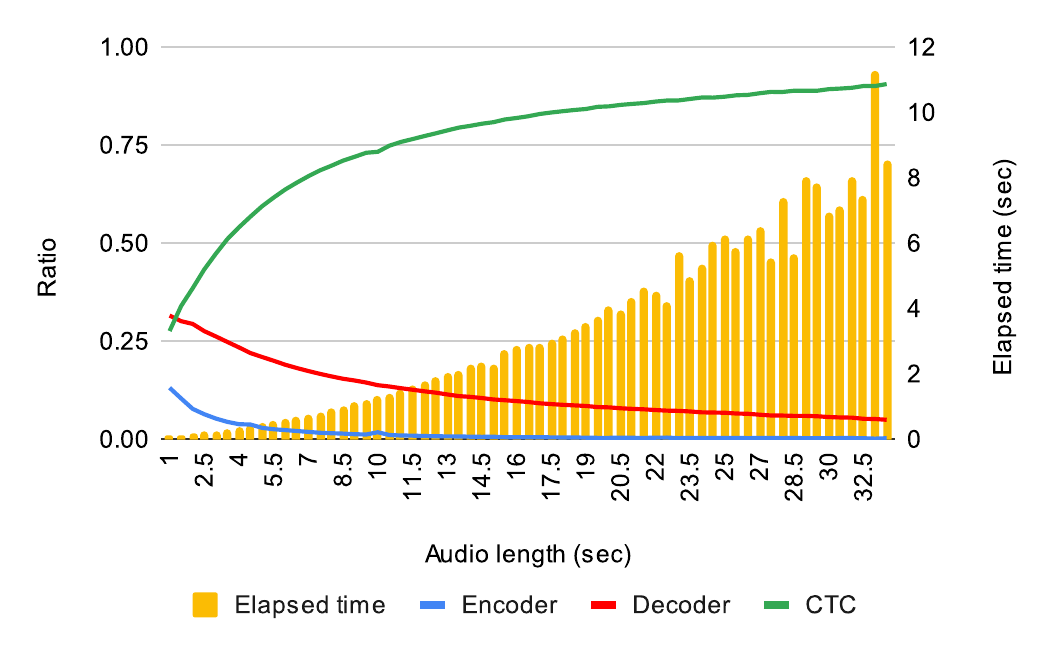}
        \subcaption{AR}
        \label{fig:decoder_ratio}
  \end{minipage} \\
  \begin{minipage}[t]{\linewidth}
    \centering
        \includegraphics[width=\textwidth]{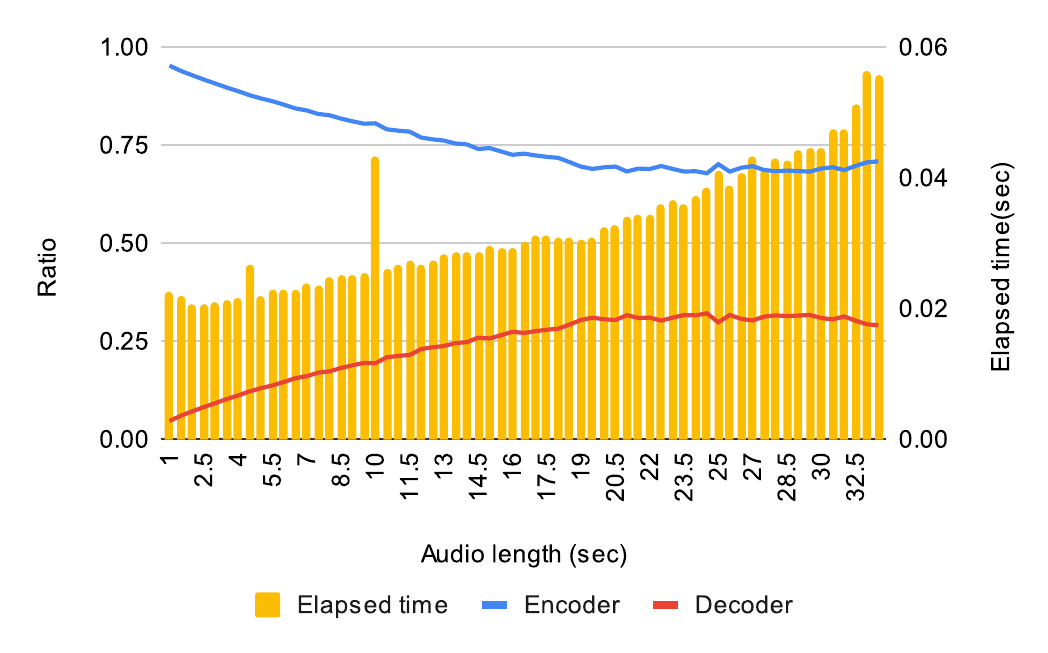}
        \subcaption{NAR}
        \label{fig:nar_decoder_ratio}
  \end{minipage}
\end{tabular}
\caption{Average inference time and proportion of time spent on the encoder, decoder, and CTC computation for (a) AR, as well as the encoder and decoder computation for (b) NAR architectures.}
\end{figure}

\section{Background}
In this section,
to provide a clear understanding of how the proposed approach combines the benefits of autoregressive (AR) and non-autoregressive (NAR) decoding,
we begin with a brief overview of the conventional encoder-decoder-based models,
including hybrid CTC-attention~\cite{watanabe2017hybrid} and Mask-CTC~\cite{higuchi20b_interspeech}.
Additionally, we investigate the role of each encoder and decoder in influencing the overall inference time for each model.

\begin{table*}[!htp]\centering
\caption{
Decoding example from the LibriSpeech test-clean set (1089-134686-0002).
The target sequence is initially predicted by gCTC and replaced with masks (``\#'').
Consecutive masks are merged into one mask.
Red tokens indicate the best hypotheses from the segment-level vectorized beam search at each iteration.
We set $\textit{max\_iteration}$ to $5$ and have five sentences for this sample.
However, we only show iterations 1 and 2 since there is no difference from the third iteration.
}
\label{tab:decoding_example}
\scriptsize
\begin{tabular}{lll}\toprule
Masked sequence &after early night\# the yellow lamps would light up here and there the squalid quarter of the\#el\# \\
iteration=1 &after early night\textcolor{red}{fall} the yellow lamps would light up here and there the squalid quarter of the \textcolor{red}{bra}el\textcolor{red}{s} \\
iteration=2 &after early nightfall the yellow lamps would light up here and there the squalid quarter of the \textcolor{red}{broth}els \\
Ground truth &after early nightfall the yellow lamps would light up here and there the squalid quarter of the brothels \\
\bottomrule
\end{tabular}
\end{table*}

\subsection{Autoregressive ASR with Hybrid CTC/Attention}
\label{sec:ar_decode}

AR decoding is a recursive method for estimating target sequences. 
In the example shown in Fig.~\ref{fig:ar_decode}, the token \verb|s| is estimated first and
then used to estimate the next token, \verb|e|.
This simple left-to-right beam search similar to \cite{seq2seqNN} is widely used in ASR to search for the most likely transcriptions~\cite{listenAttendSpell}.

However, due to the iterative nature of beam search, this can greatly slow the inference time of AR decoding. 
For example, using the hybrid CTC/attention architecture,
the beam search requires decoder computation and CTC rescoring process for each search iteration, hence inference time increases if the search process is iterated a large number of times.
This effect is shown in Fig.~\ref{fig:decoder_ratio},
which shows the proportion of time spent on the encoder and decoding process, including decoder and CTC prefix score computation,
as well as the average time spent during inference for various lengths of audio input.
This shows that as the audio length and inference time increase, the proportion of time spent in the decoding process also increases.

\subsection{Non-autoregressive ASR with Mask-CTC}
\label{sec:nar_decode}

NAR decoding is a method that avoids recursively estimating the target sequences to address the problem of slow inference.
There are various architectures for this method~\cite{Higuchi2021}, but this work focuses on the Mask-CTC~\cite{higuchi20b_interspeech} model.
In the Mask-CTC model, we first estimate the target sequences with greedy CTC decoding (gCTC) output, and then a mask is applied based on the CTC probabilities for each token.
As illustrated in Fig.~\ref{fig:nar_decode},
the token \verb|e|, \verb|a|, and \verb|m| is masked due to its lower probability.
Then, the masked token \verb|#| is estimated using the masked language model decoder.
Since the decoder is only required for the masked token, 
and the number of decoder calculations is fixed,
the number of decoder computations is significantly reduced compared to the AR method.

Fig.~\ref{fig:nar_decoder_ratio} illustrates the proportion of time spent on the encoder and decoder, and the inference time for the Mask-CTC model. 
We set the number of the decoder iterations to $10$ for measurement.
Compared to Fig.~\ref{fig:decoder_ratio}, we can see that the encoder's share of NAR is larger than AR.
Considering the number of encoder computations is always $1$, a large encoder's share means a shorter computation time.
As audio length increases, the inference time also increases, but the impact on inference time is small.
Therefore the difference in inference time between AR and NAR can be seen as the difference in the proportion of the encoder and decoding process.

However, there is an accuracy issue with the Mask-CTC-based NAR decoding.
If the number of masked tokens is different from the actual number of tokens, the accuracy degrades significantly.
For example,
if the correct sequence is \verb|s, e, a| and the masked sequence is \verb|s, #, #, a|,
the result of Mask-CTC becomes incorrect, since it tends to assign the same numbers of tokens to the masks, e.g., \verb|s, e, e, a| or \verb|s, e, a, a|.
Similarly, if the masked sequence is \verb|s, e, #, a|, it will have an insertion error.
In our preliminary experiments, we observed that almost 40\% of the masked sequences do not match the proper length.
Additionally, we have observed that insertion errors caused by the absence of a target token, occur with a probability of 2\%.

\section{Partially Autoregressive framework}

\subsection{Partially Autoregressive Inference}

To address the issues inherent in AR and NAR decoding, we propose a \textit{partially autoregressive} (PAR) decoding method.
Both AR and NAR models use the CTC and decoder components, but there is a significant difference in the trade-off balance based on usage.
AR uses an iterative process to predict the target sequence, so it does not have an accuracy issue with target sequence length.
NAR first uses gCTC results to reduce the number of tokens that need to be predicted with the decoder, resulting in fast inference.
In PAR, we combine NAR-style CTC usage and AR-style decoder usage to utilize these two advantages fully.

The architecture of PAR is illustrated in Fig.~\ref{fig:PAR_decode}.
PAR first generates a sequence of tokens by gCTC approach,
then apply the mask process using the Mask-CTC method~\cite{higuchi20b_interspeech}.
Finally, it predicts the tokens corresponding to a mask token using beam search, similarly to the AR approach.
To reduce the computational complexity, 
we propose the segment-level vectorized beam search, which significantly reduces the number of search iterations.
By applying this beam search, we can solve the NAR accuracy issue related to the incorrect target length.

As an example, let us focus on a case where only one mask is in the sequence for simplicity,
and let the $P_{\mathrm{thres}}$ represent the threshold value ranging from $0$ to $1$.
Initially, we use the gCTC result, obtained without the AR process.
However, the resulting text can contain errors due to the conditional independence assumption.
We expect these errors can be corrected by the AR process, where we utilize the pre-trained decoder and beam search.
To determine which tokens to update using the AR process, we use $P_{\mathrm{thres}}$ as a filter for posterior probability.
Tokens with probability less than $P_{\mathrm{thres}}$ will be considered less confident and replaced with the mask token.
Consecutive mask tokens are merged into a single mask because we will estimate the sequences in the AR process.
Finally, we use beam search to estimate the tokens.

We iterate the beam search for $\textit{max\_iteration}$ times for any audio length, where $\textit{max\_iteration}$ denotes the maximum number of tokens for one mask.
The value of $\textit{max\_iteration}$ is highly dependent on the threshold $P_{\mathrm{thres}}$, where a higher value is required when $P_{\mathrm{thres}}$ is closer to 1 and more tokens are replaced with mask tokens.

\begin{algorithm}
\caption{Segment-level Vectorized Beam Search}
\label{alg:s_vec_beam_search}
\begin{algorithmic}
\State Beam size: $B$ and the number of masks: $S$

\State Encoder output: $X$

\State Initialize decoder cache: $C$
\State Initialize a $S$-length list for ended hypotheses: $F_{S}$
\State Initialize hypotheses $Y_{S,B}$ and end-of-sequences $E_{S}$

\For {$i=1$ to $max\_iteration$}
    \State Calculate probability $prob = Decoder(Y_{S,B}, X, C)$
    \State Update hypotheses $Y_{S,B}$ by top-k method.

    \For {$s=1$ to $S$}
        \For {$b=1$ to $B$}
            \If {last token of $Y_{s,b}$ is $E_{s}$}
                \State $F_{s}.push(Y_{s,b})$
            \EndIf
        \EndFor
    \EndFor
\EndFor

\For {$s=1$ to $S$}
    \State Replace $s$-place mask in the masked sequence with best hypothesis in $F_{s}$
\EndFor

\end{algorithmic}
\end{algorithm}

\subsection{Segment-level Vectorized Beam Search}
\label{sec:s_vec_beam_search}

The beam search process explained in the previous section focused on the case where there is a single mask token.
However, in practice, multiple mask tokens may exist for a given sequence.
To handle multiple masks, we parallelize the decoding of masks by extending the vectorized beam search\cite{Seki2019}.
The vectorized beam search is an extension of traditional beam search that allows for the calculation of $B$ beams simultaneously.
In an offline scenario, it can also be extended to the parallel computation of $S$ utterances.
In other words, $S \times B$ hypotheses can be calculated as a single batch at each step.
In our work, we treat the value $S$ as the number of masks to enable parallel computation of all mask tokens.


The overview of the proposed segment-level vectorized beam search is described in Algorithm~\ref{alg:s_vec_beam_search}.
First, we initialize $S \times B$ hypotheses $Y_{S,B}$ from the gCTC result and masked sequence.
Initially, there is only one hypothesis, so $Y_{s,1}$ contains the gCTC result up to the corresponding mask as the hypothesis.
The rest of the hypotheses, from $Y_{s, 2}$ to $Y_{s, B}$, are initialized with dummy hypotheses.
Additionally, we store the next token for each mask in an end-of-sequence list, $E_{S}$.
Next, we calculate the probabilities of the next token for each hypothesis and update $Y_{S,B}$.
We then check, for each mask $s$, if $E_{s}$ is predicted as the next token for each of the hypothesis from $Y_{s, 1}$ to $Y_{s, B}$. We push the ended hypothesis to the list of ended hypotheses, ${F_{s}}$.
Finally, after iterating $max\_iteration$ times, we replace each mask token with the best hypothesis in $F_{s}$.
To simplify the implementation, we apply padding to the missing hypotheses so that the number of hypotheses remains constant, similar to \cite{Xiaoyu2021onnx}.

Table~\ref{tab:decoding_example} shows an example of the decoding process of a sample in the LibriSpeech $\textit{test-clean}$ set.
After calculating the gCTC result and merging the consecutive masks, there are three masks in this example.
Since the beam size $B$ is set to $10$, there are $30$ hypotheses.
In the first iteration, we correctly estimated the first and third mask tokens. 
Since we used the BPE token for this model, the red characters \verb|fall|, \verb|bra|, and \verb|s| were estimated in a single iteration.
In the second iteration, we successfully predicted the second mask, which has two corresponding tokens: \verb|bro| and \verb|th|.
This example demonstrates that it is possible to predict multiple tokens from a single mask token, enabling us to handle the accuracy issue in NAR due to incorrect target length. 
Moreover, the overall number of iterations in beam search is significantly lower than in AR by utilizing the segment-level vectorized beam search, indicating that PAR is effective in avoiding AR issues.

We determine whether to end the beam search by observing the next token of the mask token.
For example, in the example in Fig.~\ref{fig:PAR_decode},
when the first mask detects a space token \verb|_| and the second mask detects \verb|b|,
the hypothesis is terminated and pruned from the overall search. 



\begin{table*}[!htp]\centering
\caption{Dataset descriptions. The order of evaluation sets corresponds to the error results in Table~\ref{tab:experiment}. $^\dagger$Data split based on ESPnet~\cite{watanabe2018espnet}.}
\label{tab:dataset_desc}
\scalebox{0.91}{
\begin{tabular}{lcccccc}\toprule
Dataset &Language &Hours & Token & Metric &Evaluation Sets \\
\midrule
AISHELL-1~\cite{bu2017aishell} &zh &170 &char &CER &dev / test \\
JSUT~\cite{sonobe2017jsut} &ja &10 &char &CER &dev / test ($\dagger$) \\
LS-100~\cite{librispeech} &en &100 &BPE &WER &dev-\{clean,other\} / test-\{clean,other\} \\
LS-960~\cite{librispeech} &en &960 &BPE &WER &dev-\{clean,other\} / test-\{clean,other\} \\
TED-LIUM2~\cite{rousseau2014enhancing} & en &210 &BPE &WER & dev / test \\
\bottomrule
\end{tabular}
}
\end{table*}

\begin{table*}[!htp]\centering
\caption{Comparison of WER, CER, RTF, elapsed time, and memory usage for each model. RTF, inference time, and memory usage compare mean and standard deviation (in brackets) values. 
The symbols ($\downarrow$) and ($\uparrow$) indicate a lower or higher number is preferable, respectively.
}
\label{tab:experiment}
\setlength{\tabcolsep}{5pt}
\scalebox{0.86}{
\begin{tabular}{lcccccccc}\toprule
&\multicolumn{3}{c}{AR} & &\multicolumn{3}{c}{PAR} & \\
\cmidrule{2-4}\cmidrule{6-8}
Dataset &RTF ($\downarrow$) &Error [\%] ($\downarrow$) &Memory Usage [MB] & &RTF ($\downarrow$) &Error [\%] ($\downarrow$)  &Memory Usage [MB] &Speedup ($\uparrow$) \\\midrule
AISHELL-1 &0.027 (0.006) &4.6 / 5.0 &176.3 (4.2) & &0.010 (0.006) &4.6 / 4.9 &176.3 (5.9) &2.70$\times$ \\
JSUT &0.129 (0.021) &11.8 / 13.2 &208.4 (6.2) & &0.018 (0.008) &12.0 / 13.3 &199.4 (6.4) &7.17$\times$ \\
LS-100 & 0.111 (0.038) & 6.2 / 16.8 / 6.4 / 17.1 & 197.5 (25.3) & & 0.009 (0.006) & 6.5 / 17.2 / 6.7 / 17.7 & 210.6 (81.6) & 12.33$\times$ \\
LS-960 &0.110 (0.039) &2.2 / 5.2 / 2.5 / 5.2 &528.4 (44.8) & &0.008 (0.006) &2.2 / 5.5 / 2.5 / 5.6 &550.5 (132.6) &13.75$\times$ \\
TED-LIUM2 &0.182 (0.060) &7.3 / 7.1 &281.1 (47.7)& &0.012 (0.008) &7.6 / 7.3 &474.8 (298.9) &15.17$\times$ \\
\bottomrule
\end{tabular}
}
\end{table*}

\section{Experiment}

\subsection{Experimental setup}

\subsubsection{Models}
To test the effectiveness of PAR, we compare both its performance and inference speeds of various algorithms to AR and NAR.
As PAR decoding can be seen as an optimization of AR decoding, PAR can be tested using pre-trained AR models.
As a result, for both AR and PAR inference,
we used E-Branchformer models~\cite{e_branchformer} that were pre-trained on the following datasets and are publicly available on the HuggingFace hub:
AISHELL-1\footnote{https://huggingface.co/pyf98/aishell\_e\_branchformer},
JSUT\footnote{https://huggingface.co/pyf98/jsut\_e\_branchformer}
LibriSpeech-100h\footnote{https://huggingface.co/pyf98/librispeech\_100\_e\_branchformer},
LibriSpeech-960h\footnote{https://huggingface.co/asapp/e\_branchformer\_librispeech},
TED-LIUM2\footnote{https://huggingface.co/pyf98/tedlium2\_e\_branchformer}.
Note that we used two types of models trained with LibriSpeech:
a model trained with the entire dataset of $960$ hours of audio (LS-960),
and another model trained with the subset of $100$ hours of audio (LS-100).
The model trained with the LS-100 dataset is considered less accurate than the one trained with LS-960,
so we used LS-100 to investigate the effect of gCTC accuracy on PAR decoding.
For comparison against NAR decoding,
we trained models using the Conformer~\cite{conformer} architecture on LS-100.
We prepared AR and NAR models with approximately 30 million parameters for a fair evaluation.
We used the Conformer model to test if there would be any differences caused by encoder architecture.
All the models were trained and evaluated using the ESPnet toolkit~\cite{watanabe2018espnet}.

\subsubsection{Decoding setup}
For AR decoding, we set the beam size to $10$, the CTC weight to $0.3$, and employed the vectorized beam search~\cite{Seki2019}.
For PAR decoding, the beam size is set to $10$, but note that for PAR inference, we do not compute the CTC prefix score, so the CTC weight is set to $0$.
We also set the $P_{\mathrm{thres}}$ to $0.95$ and $\textit{max\_iteration}$ to $5$ for PAR decoding.
The decoding speed was measured using a single RTX 2080 Ti.

\subsubsection{Evaluation Datasets}
We used several datasets of different languages for evaluation, described in Table~\ref{tab:dataset_desc}.
For this study, we used LibriSpeech and TED-LIUM2 as English datasets, JSUT as a Japanese dataset, and AISHELL-1 as a Mandarin dataset.
For LibriSpeech, we evaluated dev and test sets, each containing $\textit{clean}$ and $\textit{other}$ sets.
Each of the five evaluation datasets was evaluated using the AR/PAR model trained on the equivalent dataset.
For example, the E-branchformer model pre-trained on the AISHELL-1 dataset was evaluated using the AISHELL-1 evaluation set.

\subsubsection{Metrics}

We evaluated accuracy with word error rate (WER) or character error rate (CER) according to the dataset.
We used the real-time factor (RTF) to measure the inference speed.
The maximum memory usage of the GPU during inference is also measured to observe the effect of mask parallel decoding.
For the RTF and memory usage, we took the mean of all evaluation sets.
The speedup column compares the RTF values with PAR compared to AR.

\begin{figure}
    \includegraphics[width=0.45\textwidth]{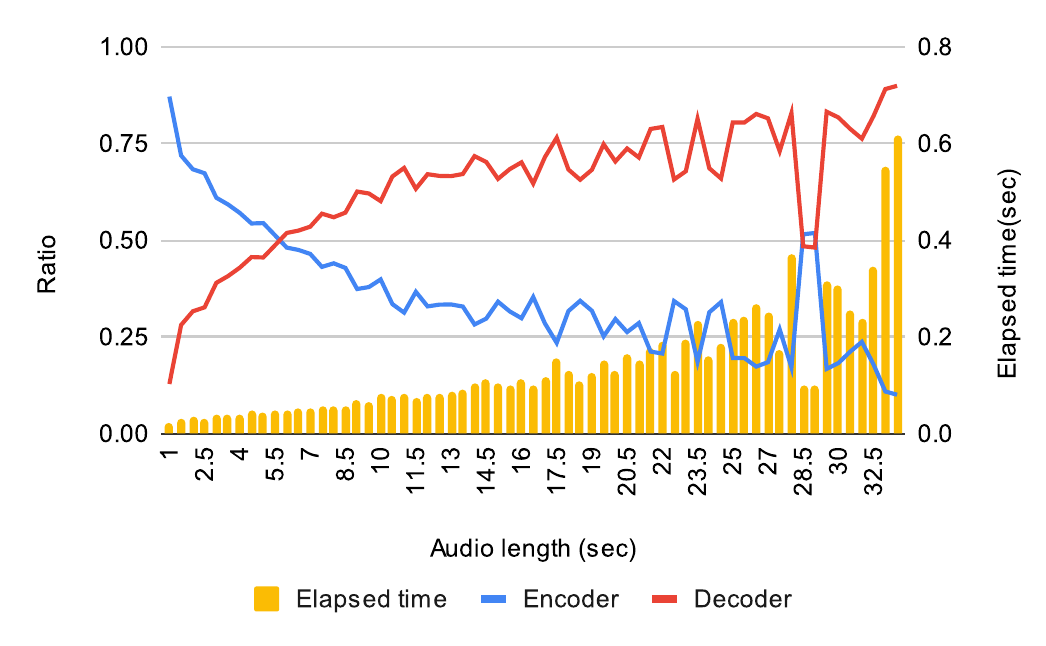}
    \caption{Average inference time and proportion of time spent on the encoder and decoder computation during PAR decoding.
    The decoder's share is greatly reduced from AR.}
    \label{fig:sar_decoder_ratio}
\end{figure}

\subsection{Results}
\label{sec:experiment_result}

\subsubsection{AR and PAR}
The evaluation results comparing AR and PAR are shown in Table~\ref{tab:experiment}.
Focusing on accuracy, it is evident that compared to AR, PAR shows a similar WER or CER for all models and evaluation datasets.
The beam search process can properly update the tokens that gCTC cannot accurately estimate.
While the accuracy has slightly degraded, this is due to the accuracy of the gCTC result.
We will explain it further in detail in the following limitations section~\ref{sec:accuracy_par}.
In terms of the RTF, we have achieved approximately $10$ times faster inference compared to AR.
In particular, since the inference speed does not largely depend on the audio length,
the standard deviation is approximately $17.5\%$ of that of AR decoding.
This feature is more effective for longer audio and in the $\textit{test-clean}$ dataset,
and we observed $89.7\times$ speed up for $29$ second audio as the maximum speedup.
Note that this speedup depends on several factors, such as the number of masks or input audio length.

Fig.~\ref{fig:sar_decoder_ratio} shows the proportion of time spent on the encoder and decoder process and the inference time for each audio length, evaluated on the LS-960 dataset.
Compared to Fig.~\ref{fig:decoder_ratio}, the decoder-to-encoder ratio is significantly reduced.
Furthermore, we can observe that the increase in inference time with respect to audio length is small, similar to that of NAR.
This is because, in this work, we set the $\textit{max\_iteration}$ to $5$, which means that the number of beam search iterations is limited to a maximum of 5.
Moreover, if there are no masks, the inference is simply a gCTC result, where we do not run decoder computation.
Therefore, the AR process does not have a high computational cost for long audio inputs.
We can confirm that the inference speed of PAR depends on the computation speed of each model, such as the encoder or decoder,
considering that the inference time slightly increases with audio length and the number of search iterations is fixed.
The computation speed of the encoder or decoder depends on the audio length, but its impact on the overall inference speed is small.

\begin{figure}
    \includegraphics[width=0.45\textwidth]{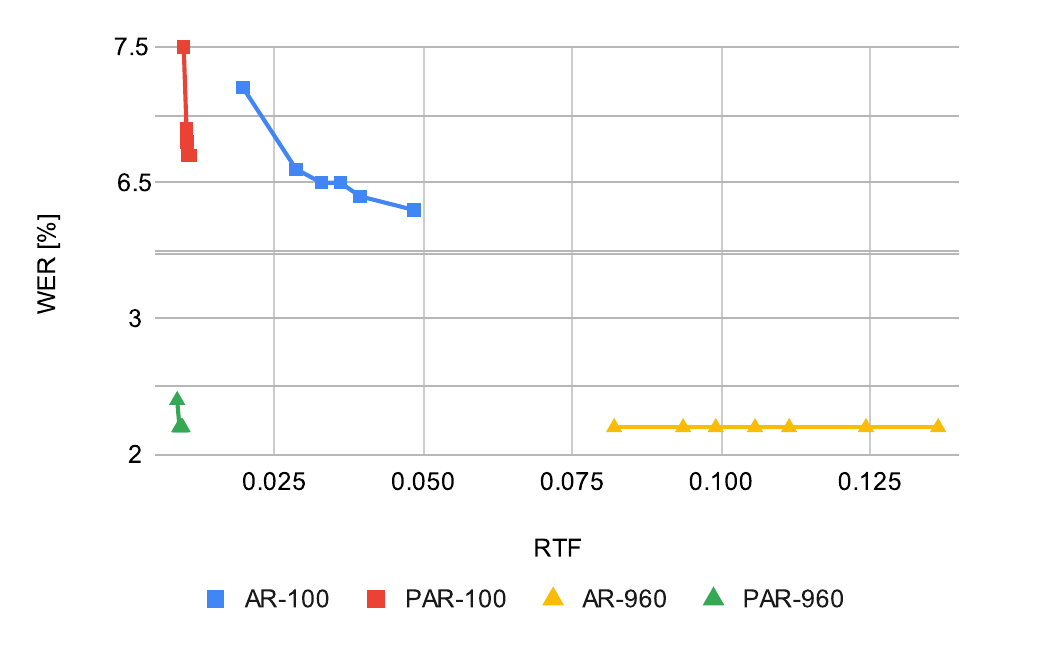}
    \caption{The comparison of WER and RTF measured using the AR and PAR methods. We used the models trained with LS-100 and LS-960 datasets and measured by changing the beam size between $1$ and $20$.}
    \label{fig:wer_rtf}
\end{figure}

We investigated the correlation between the WER and RTF as shown in Fig.~\ref{fig:wer_rtf}, by changing the beam size from $1$ to $20$, and measured the WER and RTF using the $\textit{test-clean}$ dataset from Librispeech on the E-Branchformer based pre-trained models for LS-100 and LS-960 datasets.
Comparing the AR models, the RTF differs greatly because the model sizes are different, however, the RTF of the PAR method remained almost the same.
This is because the inference time of PAR depends on the computation time of the models itself.
Therefore, the PAR curves have a significantly lower RTF compared to the AR curves, and the change in RTF is negligible as the beam-size changes, hence the narrower width.
Although the RTF remains constant, we still see the differences in WER for the PAR models, especially with the pre-trained model of the LS-100 dataset compared to the model for the LS-960 dataset.
This is due to the accuracy degradation problem we mention in Section~\ref {sec:accuracy_par}.

\subsubsection{NAR and PAR}

Comparing NAR and PAR, we can see that PAR is not as fast as NAR, as shown in Table~\ref{tab:comparison_nar}.
The number of decoder iterations is $10$, which is larger than $\textit{max\_iteration}$ for PAR,
but it performs faster than PAR.
One reason for this is that the size of the decoder input is different.
In this work, we added a dummy hypothesis as mentioned in Section~\ref{sec:s_vec_beam_search}.
As a result, the batch size of decoder input for PAR decoding is $S \times B$,
where the batch size for NAR decoding is $S$.
Therefore, it seems that the computation time per decoder is shorter for Mask-CTC NAR,
and the resulting decoder processing time becomes faster.

In terms of accuracy, we can see that PAR outperforms NAR.
Improvement in performance is similar to that of E-branchformer models in both RTF and WER, and it was confirmed that there was no difference in improvement due to architecture differences.
From these results, it is evident that applying the PAR method can solve the built-in accuracy issue for the NAR method we mentioned in Section~\ref {sec:nar_decode}.
In particular, we can confirm that PAR obtains a speed between AR and NAR but achieves similar accuracy to AR.
Therefore, a new trade-off balance that is not present in AR and NAR has been realized.

\begin{table}[!htp]\centering
\caption{Comparison with NAR. The speedup shows the speedup from the AR decoding.}
\label{tab:comparison_nar}
\scriptsize
\begin{tabular}{lccccc}\toprule
 &Model &RTF ($\downarrow$) & WER [\%] ($\downarrow$) &Speedup ($\uparrow$) \\\midrule
AR &CTC/Attention &0.198 (0.080) &6.7 / 18.3 / 7.0 / 18.6 &1.00$\times$ \\
\multirow{2}{*}{NAR} &CTC &0.005 (0.004) &7.7 / 21.0 / 7.9 / 21.4 &39.60$\times$ \\
&Mask-CTC &0.008 (0.005) &7.1 / 20.8 / 7.5 / 21.0 &24.75$\times$ \\
PAR &CTC/Attention &0.014 (0.009) &6.2 / 18.5 / 6.6 / 18.7 &14.14$\times$ \\
\bottomrule
\end{tabular}
\end{table}

\subsection{Limitations}
\label{sec:limitations}

\subsubsection{Accuracy with PAR}
\label{sec:accuracy_par}

The accuracy of PAR can be degraded if the gCTC result is not accurate.
If the result of gCTC is incorrect with high confidence,
we cannot use the AR process to refine the gCTC result.
From the comparison of the LS-100 and LS-960 models in Table~\ref{tab:experiment},
we can see that more accuracy degradation can occur with the LS-100 model.

\begin{figure}
    \includegraphics[width=0.45\textwidth]{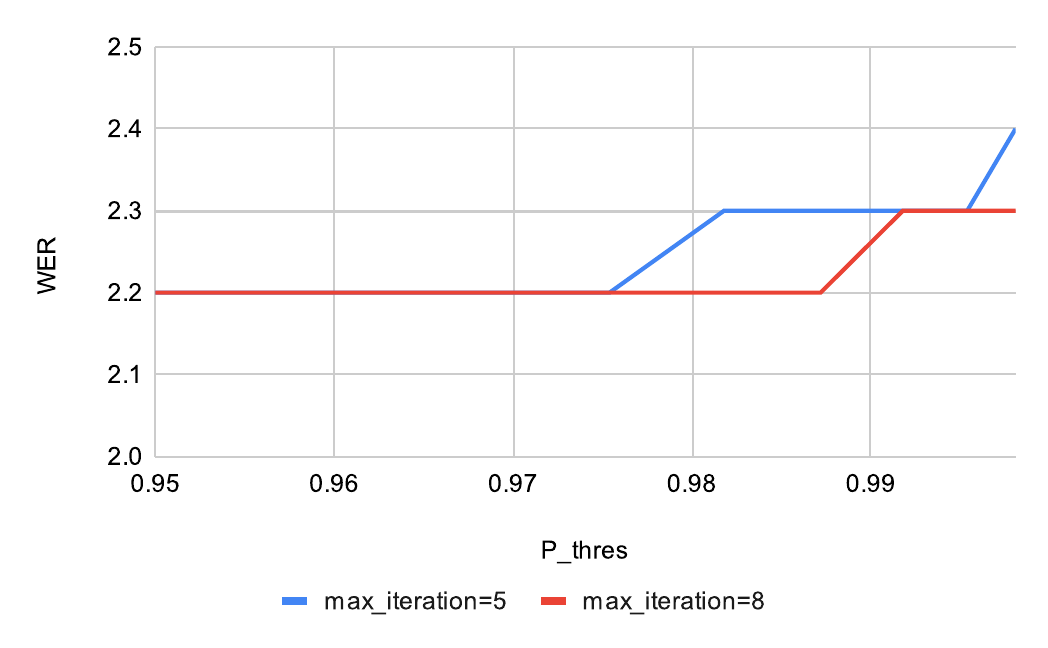}
    \caption{The relationship between the WER and $P_{\mathrm{thres}}$. We evaluated by changing the $P_{\mathrm{thres}}$ from $0.95$ to $0.999$. We used the E-Branchformer-based pre-trained model for LS-960.}
    \label{fig:wer_thresh}
\end{figure}

The accuracy may also degrade at a higher $P_{\mathrm{thres}}$ because the number of target tokens per mask may exceed $\textit{max\_iteration}$.
Since we stop the beam search after $\textit{max\_iteration}$ iterations,
if the number of target tokens exceeds $\textit{max\_iteration}$,
we cannot predict the entire sequence for one mask.
Therefore, the accuracy may be degraded if the $P_{\mathrm{thres}}$ is closer to $1.0$.
Fig.~\ref{fig:wer_thresh} describes the relationship between WER and $P_{\mathrm{thres}}$ for $\textit{max\_iteration}$.
Using a $\textit{max\_iteration}$ of $5$, we observe the accuracy degrades at higher $P_{\mathrm{thres}}$ levels.
This issue is caused by the lack of beam search iterations, so to solve this problem, we need to increase the $\textit{max\_iteration}$.
In Fig.~\ref{fig:wer_thresh}, we increased the $\textit{max\_iteration}$ to $8$ and observed a more accurate result.

\subsubsection{Memory usage}
It is important to note the standard deviation of memory usage in Table~\ref{tab:experiment} increases greatly compared to AR.
Since the decoder process in PAR is computed simultaneously for all masks, we need more GPU memory if there are many masks.
Therefore if the number of masks increases due to long audio inputs, high $P_{\mathrm{thres}}$, or low accuracy of the gCTC result, we may get an out-of-memory error as GPU memory is exceeded during inference.
Considering that the inference of all masks does not depend on each other,
it is possible to alleviate this issue by using multiple GPUs to perform inference.

\subsubsection{Segment-level Vectorized Beam Search}

If a masked sequence contains multiple masks, the predicted tokens for the second or later masks may not be accurate due to the inaccuracy of the tokens predicted for the first mask.
For example, in Fig.~\ref{fig:PAR_decode}, the masked sequence is \verb|se#_cuc#ber|, where we have two masks at the positions of tokens \verb|e| and \verb|am|.
Since we use the gCTC result to predict masked tokens, the decoder input for estimating the mask for the \verb|am| part becomes \verb|see_cuc|,
which is incorrect, instead of the correct \verb|sea_cuc|.
This incorrect decoder input might impact the accuracy of the prediction.
However, considering that even the AR decode may also have an incorrect decoder input, we believe PAR can search the same hypothesis with AR by utilizing the beam search.

\section{Conclusion}

In this work, we propose a partially autoregressive framework to obtain a new trade-off balance between accuracy and latency.
With our novel architecture design to compensate for the weakness inherent in each NAR and AR, PAR is a decoding method that takes advantage of the strengths of these two methods.
In our experiments, we observed that the AR model can be inferred at NAR-level speeds without sacrificing accuracy.
Notably, the LS-960 pre-trained model achieved a $13.75\times$ speedup with the same WER on the test-clean set.
We believe that using PAR can significantly improve the usability of the traditional hybrid CTC/Attention model.
Our framework has limitations, such as memory usage issues, that restrict the scenarios where PAR can be applied.
In future work, we plan to extend our framework to edge devices with limited computational resources.

\newpage

\bibliographystyle{IEEEbib}
\bibliography{refs}

\vspace{12pt}
\end{document}